\documentclass[12pt]{article}
\usepackage[a4paper,margin=1in]{geometry}
\usepackage{amsmath}
\usepackage{graphicx}
\usepackage{natbib}
\usepackage{url} 
\usepackage{changepage}
\usepackage[utf8x]{inputenc}
\usepackage{textcomp,marvosym}
\usepackage{fixltx2e}
\usepackage{amsmath,amssymb}
\usepackage{cite}
\usepackage{nameref,hyperref}
\usepackage{authblk}
\usepackage{algorithm}
\usepackage{algpseudocode}
\usepackage{dsfont}
\usepackage{xcolor}
\usepackage{enumitem}
\usepackage{caption,setspace}
\usepackage{pdflscape}
\usepackage{graphicx}
\usepackage{placeins}
\usepackage{mathtools}
\usepackage{appendix}

\newcommand{\dydx}[2]{\frac{\text{d} #1}{\text{d} #2}}

\renewcommand{\eqref}[1]{Equation~(\ref{#1})}

\newcommand{\CondProb}[2]{\mathbb{P}(#1 \mid #2)}

\newcommand{\PDF}[1]{p(#1)}
\newcommand{\CondPDF}[2]{p(#1 \mid #2)}

\newcommand{\like}[2]{\mathcal{L}(#1 ; #2)}
\newcommand{\loglike}[2]{\log \like{#1}{#2}}

\newcommand{\bvec}[1]{\mathbf{#1}}

\newcommand{\ind}[1]{\mathbb{I}\left(#1\right)}

\DeclareMathOperator*{\diag}{diag}

\newcommand{\paramvec}{\boldsymbol{\theta}}

\newcommand{\dat}{\mathcal{D}}
\newcommand{\simdat}{\mathcal{D}_s}

\begin{document}

\title{Mathematical modelling and uncertainty quantification for analysis of biphasic coral reef recovery patterns}


\author[1,2]{David~J. Warne\footnote{To whom correspondence should be addressed. E-mail: david.warne@qut.edu.au}}
\author[3]{Kerryn Crossman}
\author[1,2]{Grace E. M. Heron}
\author[1,2]{Jesse A. Sharp}
\author[4]{Wang Jin}
\author[1,2]{Paul Pao-Yen Wu}
\author[1,2]{Matthew J. Simpson}
\author[1,2]{Kerrie Mengersen}
\author[3]{Juan-C. Ortiz}

\affil[1]{School of Mathematical Sciences, Queensland University of Technology, Brisbane, Queensland 4001, Australia}
\affil[2]{Centre for Data Science, Queensland University of Technology, Brisbane, Queensland 4001, Australia}
\affil[3]{Australian Institute of Marine Science, Townsville, Queensland Australia}
\affil[4]{The Kirby Institute, University of New South Wales, Sydney, New South Wales, Australia}

\maketitle

\begin{abstract}
Coral reefs are increasingly subjected to major disturbances threatening the health of marine ecosystems. Substantial research is underway to develop intervention strategies that assist reefs in recovery from, and resistance to, inevitable future climate and weather extremes. To assess potential benefits of interventions, mechanistic understanding of coral reef recovery and resistance patterns is essential. Recent evidence suggests that more than half of the reefs surveyed across the Great Barrier Reef (GBR) exhibit deviations from standard recovery modelling assumptions when the initial coral cover is low ($\leq 10$\%). New modelling is necessary to account for these observed patterns to better inform management strategies.
We consider a new model for reef recovery at the coral cover scale that accounts for biphasic recovery patterns. The model is based on a multispecies Richards' growth model that includes a change point in the recovery patterns. Bayesian inference is applied for uncertainty quantification of key parameters for assessing reef health and recovery patterns. This analysis is applied to benthic survey data from the Australian Institute of Marine Science (AIMS). We demonstrate agreement between model predictions and data across every recorded recovery trajectory with at least two years of observations following disturbance events occurring between 1992--2020. This new approach will enable new insights into the biological, ecological and environmental factors that contribute to the duration and severity of biphasic coral recovery patterns across the GBR. These new insights will help to inform managements and monitoring practice to mitigate the impacts of climate change on coral reefs.  
\end{abstract}


\section{Introduction}
\label{sec:intro}

Environmental stressors, including climate change, continue to drive increased frequency and severity of disturbance events affecting coral reefs worldwide. This is leading to long-term declines in coral cover and changes in community composition~\citep{Death2009,Death2012,Duran2017,Johns2014,Lapointe2019,Tebbett2023}. To ensure the future health of coral reefs and connected marine ecosystems  we need to continue developing capabilities to effectively prioritise targeted intervention and restoration efforts~\citep{McLeod2022}. Forecasts of coral reef health under different potential environmental scenarios are an essential part of this decision-making process to ensure limited resources are efficiently allocated~\citep{Thompson2010,Vercelloni2020,Woesik2018}. In addition, predictions of expected versus observed coral cover is an important health indicator of recovery ~\citep{CastroSanguino2021,Logan2020,Thompson2020}. To produce reliable forecasts and predictions for decision-makers, we need models that are informed by our understanding of coral recovery patterns of reefs following perturbations by disturbance events, such as storm swells~\citep{Cheal2017,Death2012,Fabricius2008}, predation by crown-of-thorns starfish, \textit{Acanthaster} spp. (COTS)~\citep{Death2012,Pratchett2021,Seymour1999,Vercolloni2017}, and acute periods of thermal stress resulting in coral bleaching~\citep{Hughes2017,Mumby2021}. 

Recently, \citet{Warne2021} demonstrated evidence for biphasic recovery patterns amongst almost  $50\%$ of reefs recovering from very low cover ($\leq 10\%$), across Australia's Great Barrier Reef (GBR). Such recovery patterns were characterised by a transient period of reduced recovery that persists for approximately four years on average~\citep{Warne2021}. Such recovery patterns represent deviations from standard recovery modelling approaches that are based on variations of classical population growth models, such as exponential, Gompertz, or Logistic growth models~\citep{MacNeil2019,Ortiz2018,Simpson2022,Thompson2010,Woesik2018}. \citet{Warne2021} also demonstrate coral recovery forecasts that ignore biphasic patterns could be overestimating future median cover estimates by up to 22\% in absolute coral cover. Since the statistical methodology applied by \citet{Warne2021} is based upon multi-segment regression analysis~\citep{Aminikhanghahi2016,Hinkley1970,Jin2017}, the study was restricted to a small subset of severely disturbed reefs with at least 5 observations of recovery. This motivates the development of a recovery modelling framework that can account for the effects of biphasic recovery in a broader range of scenarios. 

In the biological and ecological modelling literature, biphasic models have been widely studied and applied in a variety of settings~\citep{Bodnar2013,Brouwer2017,Honsey2016,A.Moeller2021,Phaiboun2015}, the most general of which has been demonstrated recently by \citet{Murphy2022}. Here, we present a multi-species biphasic coral recovery model building on the modelling frameworks of~\citet{Murphy2022} and \citet{Simpson2023}. Using Bayesian analysis for model calibration, prediction, and uncertainty quantification, we demonstrate the approach is directly applicable to inform health assessment of the reefs within the GBR based upon coral cover monitoring data~\citep{Jonker2008}. Through embedding our recovery model within a Bayesian analysis framework, we provide a powerful tool for uncertainty quantification of key recovery parameters and coral cover predictions that account for this parameter uncertainty.

\section{Materials and methods}
\label{sec:methods}
In this section we briefly describe the benthic survey data used in this work. A description of our mathematical model that captures biphasic recovery patterns is then presented along with the Bayesian methods for model calibration and assessment of model fitness.

\subsection{Benthic survey data}
\label{sec:materials}
\begin{figure}[h]
	\centering
	\includegraphics[width=\linewidth]{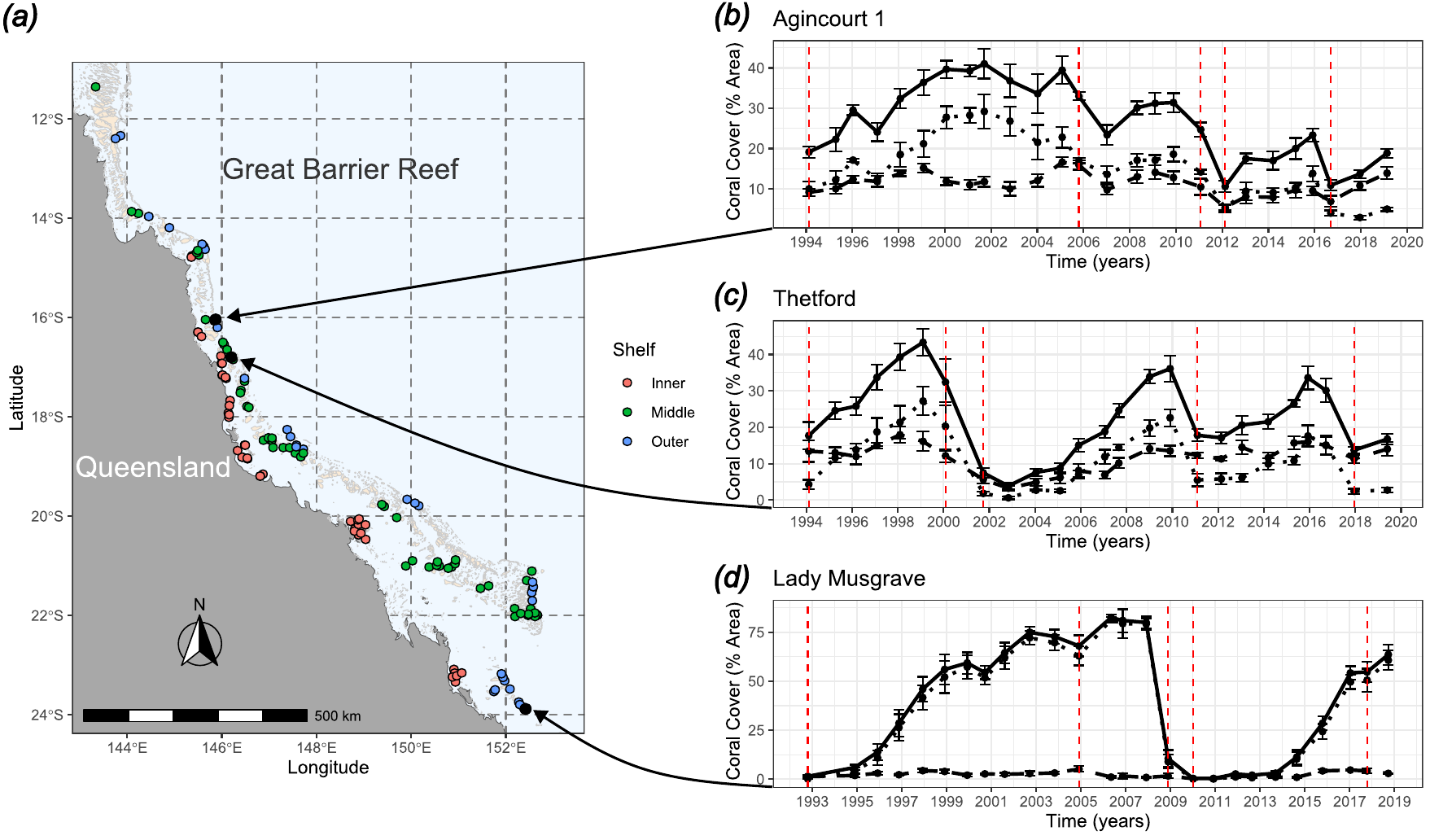}
	\caption{(a) Location of reef sites that are monitored as part of the Long-Term Monitoring Program (LTMP) and the Marine Monitoring Program (MMP). The MMP monitors the inshore reefs (red), and the LTMP monitors the Middle (green) and Outer (blue) reef shelves. Example coral cover time-series are shown for (b) Agincourt 1, (c) Thetford, and (d) Lady Musgrave reefs with their locations indicated with arrows (black) in (A). Here, total hard coral cover (black solid), the fast growing \textit{Acropridae} family (black dotted), and other hard corals (black dashed) are shown over time along with official disturbance event records (red dashed). Error bars indicate the standard error of the coral cover record based on the variation at the transect level.}
	\label{fig:GBR}
\end{figure}

Coral cover time-series data are obtained from benthic surveys undertaken by the Australian Institute of Marine Sciences (AIMS) Long Term Monitoring Program (LTMP) and Marine Monitoring Program (MMP) between 1992 to 2020. Combined, these collection programs include taxonomically classified cover data obtained from over 373 sites across 135 reefs. These monitored reefs  are characteristic of the inshore, middle and outer shelf areas of the GBR (Figure \ref{fig:GBR}(a)). The dataset includes a mixture of annual and biennial surveys at depths between 2m-9m. At each site, five permanent transects (dimension 5m$\times$50m for LTMP and 5m$\times$20m for MMP) are photographed at regular intervals (1m intervals for LTMP and 0.5m intervals for MMP). Coral cover is estimated and classified using a 5-point stencil overlaid in each image (See~\citet{Jonker2008} for details on the data collection protocol). Example coral cover time series are shown for Agincourt 1 (Figure \ref{fig:GBR}(b)), Thetford (Figure \ref{fig:GBR}(c)) and Lady Musgrave reefs (Figure \ref{fig:GBR}(d)).

Following \citet{Ortiz2018}, and \citet{Warne2021}, disturbances were identified by statistically significant decline ($p \leq 0.05$) in cover according to a one-sided paired t-test at the transect level ($n= 5$) for consecutive visits (Figure~\ref{fig:GBR}B--D). A recovery trajectory is defined as the sequence of monitoring surveys between consecutive disturbances inclusive of the disturbance event defining the start of the recovery. Across the GBR, the patterns of coral recovery between disturbance events varies widely, as shown in the examples in Figure~\ref{fig:GBR}. This motivates a modelling approach that can account for deviations from standard growth patterns typically defined using a Gompertz or logistic model~\citep{Thompson2015,Thompson2020,Simpson2023}. One limitation of the methodology developed in \citet{Warne2021} was the restriction to recovery trajectories with low initial cover ($\leq 10 \%$) and have $n \geq 5$ benthic survey visits before the next disturbance. Our new modelling framework reduces these restrictions, requiring only that the number of visits between subsequent disturbances is $n \geq 3$ with no restrictions on the initial coral cover are necessary. From the LTMP and MMP benthic surveys, we obtain $N = 737$ recovery trajectories among $336$ reef sites and $124$ reefs across the GBR. 

Throughout this manuscript, we denote $\bvec{X}^{(i)}(t_j)$ as the vector of taxonomically classified coral cover data at the reef site level for the $i$-th recovery trajectory, $i = 1,2,\ldots, N$, as surveyed at time $t_j$, and $j = 0,1,\ldots, n^{(i)}$. Here, $j=0$ relates to the initial visit in which the reef site was perturbed by a disturbance event and $n^{(i)}$ is the number of subsequent visits in the $i$-th recovery trajectory. Thus the $i$-th observed recovery trajectory is $\dat^{(i)}=[\bvec{X}^{(i)}(t_0),\bvec{X}^{(i)}(t_1),\ldots,\bvec{X}^{(i)}(t_{n^{(i)}})]$ containing the time-series of benthic survey data for that recovery trajectory.

\subsection{Model development}

Our mathematical model of coral reef recovery accounts for biphasic recovery patterns as identified in previous work \citep{Warne2021}, and allows for more general single-phase recovery patterns beyond standard approaches such as logistic or Gompertz growth models~\citep{Simpson2022, Thompson2010,Thompson2020}. We first describe a single species version of the model, before extending this to a multi-species model that also accounts for interactions and competition between coral types.

\subsubsection{Generalised logistic growth}
\begin{figure}[h]
	\centering
	\includegraphics[width=0.9\linewidth]{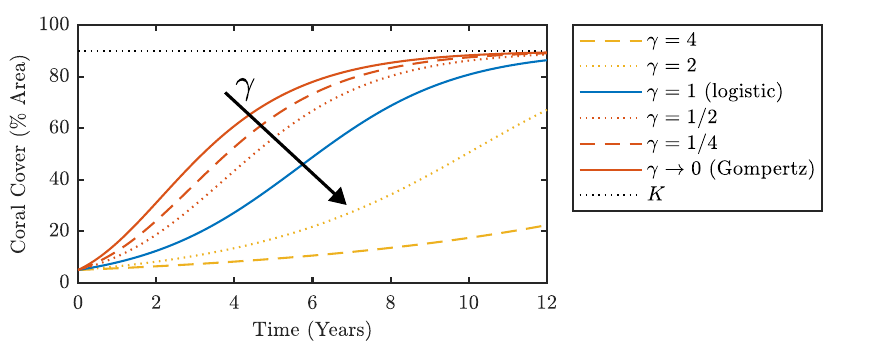}
	\caption{The effect of the generalisation parameter, $\gamma$, in Richards' growth model. The logistic growth model corresponds to $\gamma = 1$ (blue solid). For $\gamma \to 0$ we obtain the Gompertz growth model with rapid initial recovery and stronger effect of spatial competition. Other values for $\gamma$ result is alternative recovery models. Here, $K = 90 \%$, $\alpha = 0.5\,(\text{Years}^{-1})$, and $c_0 = 5 \%$.}
	\label{fig:richardsgrowth}
\end{figure}

We begin with the Richards' growth model \citep{Richards1959,Tjorve2010,Simpson2022}, also known as the \textit{generalised logistic growth model}, given by 
\begin{equation}
\underbrace{\dydx{C(t)}{t}}_{\text{Rate of change in cover}} = \underbrace{\alpha C(t)}_{\text{Coral growth}} \times \underbrace{\frac{1}{\gamma}\left[1 - \left( \frac{C(t)}{K} \right)^\gamma \right]}_{\text{Competition for available space}}, \quad t > t_0
\label{eq:glgm}
\end{equation} 
where $K \in (0,100]$ (\% of area) is the maximum \% of area the coral can feasibly cover, i.e., the available space,  $C(t) \in (0,100]$ (\% of area) is the hard coral cover at time $t$ (years), $\alpha > 0$ (1/years) is the intrinsic rate of cover increase and $\gamma> 0$ is a non-dimensional generalisation parameter that controls the shape of the sigmoid growth curve. For example, logistic growth is recovered for $\gamma = 1$ and Gompertz growth occurs in the limit as $\gamma \to 0$. The effect of $\gamma$ on the recovery curve is shown in Figure~\ref{fig:richardsgrowth}. Note that Richards' growth model is usually presented without the $1/\gamma$ factor as it can be absorbed into the definition of the rate parameter, $\alpha$. However, \citet{Simpson2022} show that this leads to non-identifiability in the rate parameter and furthermore renders comparison of $\alpha$ across different values of $\gamma$ to be meaningless.

\FloatBarrier
\subsubsection{Incorporating biphasic recovery}
\label{sec:singlespecies}
To incorporate biphasic growth into the model, we build on recent developments by \citet{Murphy2022} who consider a piecewise defined population dynamic model,
\begin{equation}
\dydx{C(t)}{t} = \begin{cases} f_1(C(t)) & \text{if } t \leq T\\
f_2(C(t)) & \text{if } t > T
\end{cases},
\label{eq:bpgm}
\end{equation}
where the rate of change of a population, $C(t)$, switches from $f_1(C(t))$ to $f_2(C(t))$ at the \textit{change point} $t = T$. It is important to note that the solution to \eqref{eq:bpgm} is still continuous at this change point despite the derivative potentially being discontinuous. 

To mathematically model the biphasic recovery patterns identified by~\citet{Warne2021}, we require $f_1(C(t)) < f_2(C(t))$ that is for the time $t \in (t_0,T]$ the coral recovery rate parameter is lower than when $t \in (T, \infty)$. We achieve this through defining $f_1(C(t)) = \alpha_d f_2(C(t))$ where  $\alpha_d \in (0,1)$ is the recovery scale factor, and $f_2(C(t))$ corresponds to the Richards' growth model (\eqref{eq:glgm}). We then define a duration, $T_d > 0$, after which the recovery reverts back to \eqref{eq:glgm}. Throughout, we refer to this time $T_d$ as the \textit{relative change point}, since it is duration of the first phase relative to the initial time $t_0$, that is $T = T_d + t_0$. Substituting \eqref{eq:glgm} into the general biphasic model of~\citet{Murphy2022} (\eqref{eq:bpgm}) leads to the following  biphasic Richards' growth model,
\begin{equation}
\dydx{C(t)}{t} = \begin{cases} \dfrac{\alpha_d\alpha}{\gamma} C(t)\left[1 - \left( \dfrac{C(t)}{K} \right)^\gamma \right] & \text{if } t_0 < t \leq t_0 + T_d \\[1em]
\dfrac{\alpha}{\gamma} C(t)\left[1 - \left( \dfrac{C(t)}{K} \right)^\gamma \right] & \text{if } t > t_0 + T_d
\end{cases}.
\label{eq:2pglgm}
\end{equation}
 Given a known initial condition $C(t_0)=c_0$ and values for the parameters $\paramvec = [\alpha, \gamma, T_d,\alpha_d, K]$, \eqref{eq:2pglgm} admits the analytical solution, 
\begin{equation}
C(t) = \begin{cases}
K\left\{1 + [(K/C_0)^\gamma-1]\exp{\left(-\alpha_d\alpha(t - t_0)\right)}\right\}^{-1/\gamma} & \text{if } t_0 < t \leq t_0 + T_d\\
K\left\{1 + [(K/C_d)^\gamma-1]\exp{\left(-\alpha(t - T_d)\right)}\right\}^{-1/\gamma} & \text{if } t > t_0 + T_d
\end{cases},
\label{eq:2pglgmsol}
\end{equation}
where $C_d = C(t_0+T_d) = K\left\{1 + [(K/C_0)^\gamma-1]\exp{\left(-\alpha_d\alpha T_d\right)}\right\}^{-1/\gamma}$. 

\subsubsection{Accounting for multiple coral family groups}
\label{sec:multispecies}
Due to the substantial differences in the recovery rate of the \textit{Acroporidae} family of corals, it is common to model their recovery distinctly from other hard coral families~\citep{Thompson2010,Thompson2020,Warne2021}. We consider the joint recovery of $M$ coral family groupings with $C_m(t)$ denoting the coral cover of family group $m \in [1,2,\ldots,M]$. This leads to a coupled system of $M$ biphasic Richards' growth model (\eqref{eq:2pglgm}) with an shared maximum coral cover term to account for competition for space between species,
\begin{equation}
\dydx{C_m(t)}{t} = \begin{cases} \dfrac{\alpha_{d,m}\alpha_m}{\gamma_m} C_m(t)\left[1 - \left( \dfrac{\sum_{k=1}^M C_k(t)}{K} \right)^{\gamma_m} \right] & \text{if } t_0 < t \leq t_0 + T_{d,m} \\[1em]
\dfrac{\alpha_m}{\gamma_m} C_m(t)\left[1 - \left( \dfrac{\sum_{k=1}^M C_k(t)}{K} \right)^{\gamma_m} \right] & \text{if } t > t_0 + T_{d,m}
\end{cases},
\label{eq:2pglgmms}
\end{equation} 
where $m = 1,2,\ldots,M$ and $\paramvec = [\alpha_1, \gamma_1, T_{d,1},\alpha_{d,1},\alpha_2, \gamma_2, T_{d,2},\alpha_{d,2},\ldots,\alpha_M, \gamma_M, T_{d,M},\alpha_{d,M}, K]$. For this work, we only consider the case of $M = 1$ or $M = 2$. When $M=1$, we consider only the total hard coral cover, and use the analytical solution given in \eqref{eq:2pglgmsol}. When $M=2$, we use $C_1(t)$ to denote the coral cover of the \textit{Acroporidae} family and $C_2(t)$ to denote the coral cover of all other hard corals. See Section~\ref{sec:discuss}, \citet{Thompson2010}, and \citet{Thompson2020} for discussion around this choice of splitting.

The analytic solution (\eqref{eq:2pglgmsol}) for the single species model (\eqref{eq:2pglgm}) does not extend to the multispecies case (\eqref{eq:2pglgmms}). Therefore, we rely on numerical schemes to solve the coupled system. We apply the Runge-Kutta-Fehlberg method with adaptive time-stepping \citep{Fehlberg1969} as implemented in the \texttt{R} package, \texttt{deSolve}~\citep{Soetaert2010}. 

\subsubsection{The observation process model}
Let $\bvec{C}(t) = [C_1,C_2,\ldots, C_M(t)]^\text{T}$ be the vector of coral cover for each of the $M$ modelled family groups. We then model observation noise using a Gaussian distribution, that is the observed cover, $\bvec{X}(t)$,  at time $t$ is given by
\begin{equation}
\bvec{X}(t) \sim \mathcal{N}(\bvec{C}(t),\boldsymbol{\Sigma}(t)),
\label{eq:obsproc}
\end{equation}
where $\boldsymbol{\Sigma}(t) = \diag{\{\hat{s}_m(t)^2/5\}}$ is a diagonal matrix of standard error estimates where $\hat{s}_m(t)^2$ is the empirical variance in coral cover for family group $m$ at time $t$ taken across the five transects (See Section~\ref{sec:materials}). Alternative observation processes could be considered, such as binomial noise, however, the observed heteroscedasticity in the cover data tends to be different than expected for multinomial noise models, that is, the data demonstrates both overdispersion and underdispersion. Using a Gaussian noise approximation allows for a more flexible treatment of the variability that is also computationally efficient.

\subsection{Bayesian uncertainty quantification}
\label{sec:bayesinf}
Inference is performed within a Bayesian framework, that is, for each recovery trajectory $i \in [1,2,\ldots,N]$ we require samples from the joint posterior distribution~\citep{Gelman2013} with density,
\begin{equation*}
\CondPDF{\paramvec^{(i)}}{\dat^{(i)}} = \frac{\like{\paramvec^{(i)}}{\dat^{(i)}}\PDF{\paramvec^{(i)}}} {\PDF{\dat^{(i)}}},
\end{equation*}
where $\like{\paramvec^{(i)}}{\dat^{(i)}}$ is the likelihood of the observed benthic survey data, $\dat^{(i)}$, under the model given parameters, $\paramvec^{(i)}$, $\PDF{\paramvec^{(i)}}$ is the prior probability density and $\PDF{\dat^{(i)}}$ is the evidence, that is, the marginal likelihood given by
\begin{equation*}
	\PDF{\dat^{(i)}} = \int \like{\paramvec^{(i)}}{\dat^{(i)}}\PDF{\paramvec^{(i)}}\textrm{d}\paramvec^{(i)}.
\end{equation*} 
Conditional on the model solution $\bvec{C}(t_j; \paramvec^{(i)})$ (which will be  numerical for $M > 1$) we obtain the log-likelihood as a multivariate Gaussian around the model,
\begin{equation}
 \loglike{\paramvec^{(i)}}{\dat^{(i)}}= -\frac{1}{2}\left\{n^{(i)}\log(2\pi) +\sum_{j=0}^{n^{(i)}} \left[\log{|\boldsymbol{\Sigma}^{(i)}(t_j)|} + \Delta\bvec{C}(t_j; \paramvec^{(i)})^{\text{T}}\boldsymbol{\Sigma}^{(i)}(t_j)^{-1}\Delta\bvec{C}(t_j; \paramvec^{(i)})\right] \right\},
\label{eq:loglike}
\end{equation}
where $\Delta\bvec{C}(t_j; \paramvec^{(i)}) = \bvec{X}^{(i)}(t_j) - \bvec{C}(t_j; \paramvec^{(i)})$ is the difference between the observations and modelled mean recovery at time $t_j$ with $j = 0,1,\ldots, n^{(i)}$. 

We assume independent priors, that is,
\begin{equation}
\PDF{\paramvec^{(i)}} = \prod_{m=1}^M \PDF{\alpha_m^{(i)}}\PDF{\gamma_m^{(i)}}\PDF{T_{d,m}^{(i)}}\PDF{\alpha_{d,m}^{(i)}}.
\label{eq:prior}
\end{equation}
Note that the maximum coral cover, $K^{(i)}$, is not inferred since it is standard practice to estimate this directly from the LMTP and MMP Data~\citep{Thompson2020} (See Section~\ref{sec:discuss} for discussion). For the remaining parameters we have, $\alpha_m^{(i)} \sim \mathcal{U}(0,1)$, $\alpha_{d,m}^{(i)} \sim \mathcal{U}(0,0.9)$, $\gamma_m^{(i)} \sim \mathcal{U}(0,2)$, $T_{d,m}^{(i)} \sim \mathcal{U}(0,D^{(i)})$
where $D^{(i)} = t_n^{(i)} - t_0^{(i)}$ is the duration of the $i$th recovery trajectory and   $\mathcal{U}(a,b)$ denotes a uniform distribution over the interval $[a,b]$. By using uniform priors, we aim for the priors to be vague or uninformative within the physically feasible ranges, however, we note that general guarantees of uninformativity are not practically available~\citep{Efron1986,Warne2019}. 

For our model, the evidence term is intractable. As a result, we can only evaluate the posterior density up to a normalising constant using \eqref{eq:loglike} and \eqref{eq:prior}. To deal with this, we apply Markov chain Monte Carlo (MCMC). In our case, the commonly applied sampler available in the \texttt{Stan} environment~\citep{Carpenter2017} or other samplers based on Hamiltonian Monte Carlo~\citep{Duane1987} are not appropriate due to a discontinuity that arises in $\nabla \loglike{\paramvec^{(i)}}{\dat^{(i)}}$ from the change-points $T_{d,m}$. While numerical smoothing of the discontinuities is feasible, we avoid this by selecting a more general MCMC sampler that does not require $\loglike{\paramvec^{(i)}}{\dat^{(i)}}$ to be differentiable. Specifically, we use the robust adaptive random-walk Metropolis-Hastings MCMC scheme of \citet{Vihola2012} that is implemented within the \texttt{R} package, \texttt{adaptMCMC} \citep{Scheidegger2021}. 

For each trajectory, $\dat^{(1)}, \dat^{(2)},\ldots,\dat^{(N)}$, we use four independent chains to assess standard convergence diagnostics of effective sample size (ESS) and the Gelman $\hat{R}$ statistic~\citep{Gelman1992,Vehtari2020} using the \texttt{coda} package \citep{Plummer2006}. Sampling is terminated when $\hat{R} \leq 1.1$ and  $\text{ESS} \geq 200$ for each parameter. This results in $N = 737$ sets of posterior samples. Thinning is applied for chains with very large $\text{ESS}$ to save memory requirements while still ensuring the convergence criteria are satisfied.

To assess the model fit, we sample from the posterior predictive distribution with density,
\begin{equation}
\CondPDF{\simdat^{(i)}}{\dat^{(i)}} = \int \CondPDF{\simdat^{(i)}}{\paramvec^{(i)}}\CondPDF{\paramvec^{(i)}}{\dat^{(i)}}\, \text{d}\paramvec^{(i)},
\label{eq:ppd}
\end{equation}
where $\simdat^{(i)}$ is simulated trajectory data using the biphasic recovery model (\eqref{eq:2pglgmms}) and observation process (\eqref{eq:obsproc}) given a random draw from the posterior samples as obtained via MCMC. Using these simulations, we obtain a probability density estimate for the model uncertainty accounting for all of the underlying parameter uncertainty. This density is then used to estimate 50\%, 90\%, 95\%, and 99\% credible intervals (CrI). This enables the model fit to be assessed by inspecting the location of the observed data within these reported intervals.

\section{Results}
\label{sec:results}
\begin{figure}[ht]
	\centering
	\includegraphics[width=0.75\linewidth]{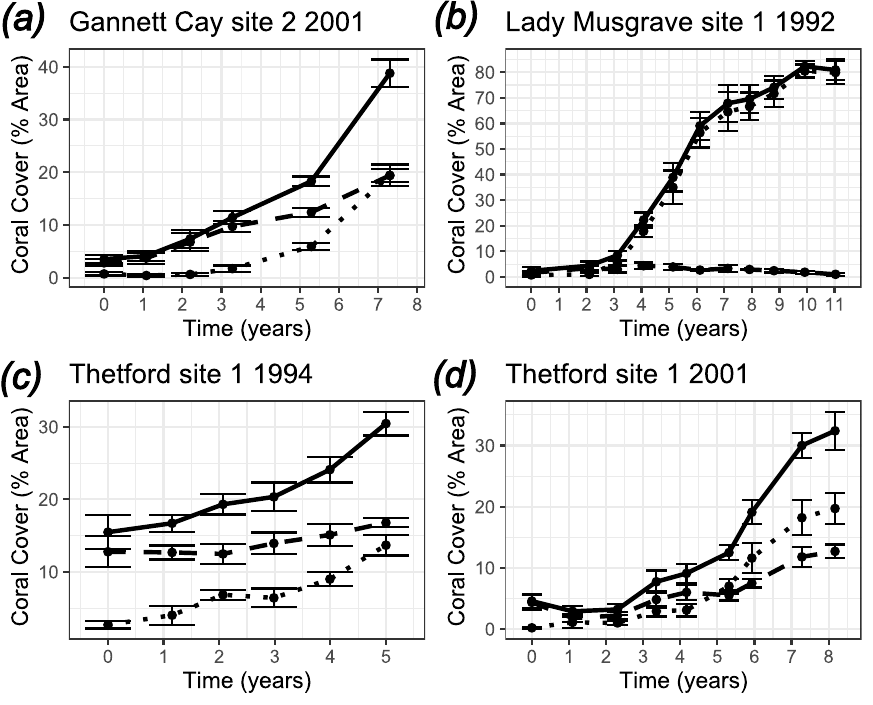}
	\caption{Example recovery trajectory data used to highlight the efficacy of our model in capturing different recovery patterns observed for hard coral (solid black lines) broken down into the \textit{Acroporidae} (dotted black lines) and all other hard corals (dashed black lines). Error bars indicate standard errors for the coral cover estimates. }
	\label{fig:examplerecovery}
\end{figure}

We calibrate the biphasic generalised recovery model using Bayesian analysis for the $N =737$ recovery trajectories as described Section~\ref{sec:methods}. The resulting output is posterior distributions and posterior predictive samples for each of the $N=737$ recovery trajectories. In this section, we provide summaries of the overall model fitness in terms of convergence diagnostics and posterior predictive probabilities. We also present example outputs using four  representative recovery trajectories (Figure~\ref{fig:examplerecovery}): site 2 of Gannett Cay reef between 2001--2009 (Figure~\ref{fig:examplerecovery}(a)), site 1 of Lady Musgrave reef between 1992--2003 (Figure~\ref{fig:examplerecovery}(b)), and site 1 of Thetford reef between 1994--1999 (Figure~\ref{fig:examplerecovery}(c)) and between 2001--2009 (Figure~\ref{fig:examplerecovery}(d)). For ease of description we refer to these recovery trajectories by their reef name and start year, that is Gannett Cay 2001 (Figure~\ref{fig:examplerecovery}(a)), Lady Musgrave 1992 (Figure~\ref{fig:examplerecovery}(b)), Thetford 1994 (Figure~\ref{fig:examplerecovery}(c)), and Thetford 2001 (Figure~\ref{fig:examplerecovery}(d)). 

These example recovery trajectories are selected to demonstrate the range of bisphasic recovery patterns that are observed across the GBR while also providing interesting interplay between the single species and two species models.

\subsection{MCMC convergence diagnostics}
\label{sec:convdiag}

For each recovery trajectory we obtain samples from the joint posterior distribution under both the single-species (\eqref{eq:2pglgm}) and the two-species (\eqref{eq:2pglgmms}) versions of the biphasic Richards' growth model. For the single species model we infer the intrinsic growth rate $\alpha$, the shape parameter, $\gamma$, the relative change-point, $T_d$, and the scale factor, $\alpha_d$. For the two-species model we infer equivalent parameters for the \textit{Acroporidae} family, $\alpha_A, \gamma_A, T_{d,A}, \alpha_{d,A}$, for all other hard corals, $\alpha_C,\gamma_C, T_{d,C},\alpha_{d,C}$. In both cases, we used standard methods to estimate maximum coral cover for reefs based on abiotic cover and transient silt records~\citep{Jonker2008,Thompson2010,Thompson2020}.  

As described in Section~\ref{sec:bayesinf} we terminate sampling when four independent chains achieve $\widehat{R} \leq 1.1$ and ESS $\geq 200$. Of the $N = 737$ recovery trajectories, $38$ did not meet this convergence criteria after 200,000 MCMC iterations. Most of these failed convergences where due to ESS $< 200$, with only 11 failing due to $\hat{R} > 1.1$. The lower ESS indicates slower mixing in the MCMC chains and a likely cause is the parameter non-identifiability issues that arises when there a very weak biphasic recovery pattern leading to a trade-off between $T_d$, $\alpha_d$ and $\gamma$, that is the data does not have enough information to distinguish between a single-phase model with $T_d = 0$ and a biphasic model $T_d > 0$. In the very few cases where $\hat{R} > 1.1$, it seems that the proposed model is a poor fit, and we discuss potential improvements to account for these cases in the discussion (Section~\ref{sec:discuss}). That is, we are able to reliably characterise model parameters for $95\%$ of the available recovery trajectories ($N = 699$) with only $1.5\%$ ($N=11$) failing due to serious convergence issues arising from poor model fit.  
\begin{figure}[ht]
	\centering
	\includegraphics[width=\linewidth]{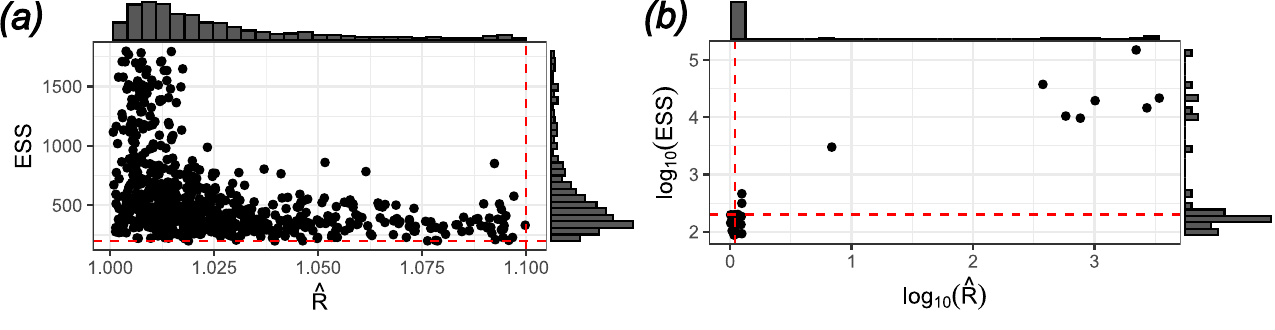}
	\caption{Scatter plots summarising the MCMC convergence diagnostics for each recovery trajectory after a maximum of 200,000 iterations per chain  and four independent chains. Convergence thresholds are indicated: $\hat{R} \leq 1.1$ (vertical red dashed line), $\text{ESS} \geq 200$ (horizontal red dashed line). (a) For the majority of recovery trajectories, the MCMC diagnostics comfortably meet the thresholds. (b) Of the small number of recovery trajectories that fail in their MCMC convergence diagnostics, most are close to the threshold.  }
	\label{fig:convsum}
\end{figure}

\begin{figure}[ht]
	\centering
	\includegraphics[width=\linewidth]{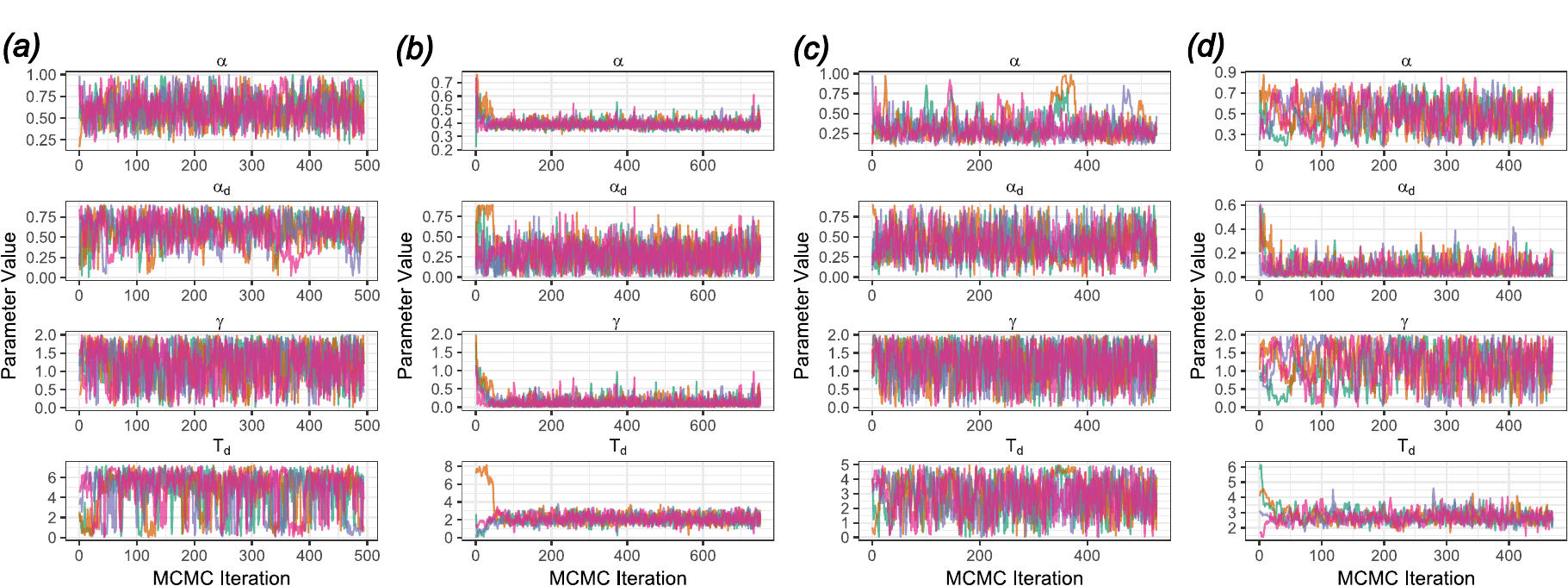}
	\caption{Trace plots showing good mixing and agreement between four independent MCMC chains (green, orange, purple, and magenta solid lines) in all parameters. (a) Gannett Cay 2001, (b) Lady Musgrave 1992, (c) Thetford 1994, and (d) Thetford 2001.}
	\label{fig:exampletrace}
\end{figure}

\begin{table}[h]
		\centering
	\caption{Convergence diagnostic results for the four example recovery trajectories.}
\begin{tabular}{l|rrr}
	\hline
	Recovery Trajectory & Parameter & $\hat{R}$ [Upper $95\%$CI] & ESS\\
	\hline
	\hline
	  & $\alpha$   & $1.02$ $[1.06]$ & 429\\
	Gannett Cay site 2 2001	 & $\alpha_d$ & $1.01$ $[1.02]$ & 1092\\
							 & $\gamma$   & $1.01$ $[1.04]$ & 990\\
							 & $T_d$      & $1.02$ $[1.05]$ & 227\\
							 \hline
	  & $\alpha$   & $1.00$ $[1.00]$ & 1074\\
	Lady Musgrave site 1 1992 & $\alpha_d$ & $1.00$ $[1.01]$ & 707\\
	& $\gamma$   & $1.00$ $[1.01]$ & 446\\
	& $T_d$      & $1.00$ $[1.01]$ & 454\\
	\hline
	 & $\alpha$   & $1.00$ $[1.01]$ & 969\\
	Thetford site 1 1994& $\alpha_d$ & $1.08$ $[1.15]$ & 287\\
	& $\gamma$   & $1.00$ $[1.02]$ & 1549\\
	& $T_d$      & $1.01$ $[1.03]$ & 589\\
	\hline
	 & $\alpha$   & $1.01$ $[1.01]$ & 576\\
	Thetford site 1 2001& $\alpha_d$ & $1.01$ $[1.03]$ & 395\\
	& $\gamma$   & $1.01$ $[1.04]$ & 365\\
	& $T_d$      & $1.01$ $[1.03]$ & 402\\
	\hline 
\end{tabular}
\label{tab:conv}
\end{table}

In the context of the four examples trajectories (Figure~\ref{fig:examplerecovery}), the convergence diagnostics in the form of MCMC trace plots, parameter-wise ESS and $\widehat{R}$  statistics are shown in Figure~\ref{fig:exampletrace} and Table~\ref{tab:conv} for the single species model. The trace plots show good mixing and agreement across all four independent MCMC chains (Figure~\ref{fig:exampletrace}). In addition the estimated $\widehat{R}$  statistics including the upper $95\%$CI are well below the standard threshold of $1.1$ (with exception of the $\widehat{R}$ upper $95\%$CI for $\alpha_d$ at Thetford 1994), and the ESS for all parameters exceed 200. Given these results, it is reasonable to conclude our convergence criteria are appropriate and that the Markov chains have reached stationarity. 
\FloatBarrier

\subsection{Assessment of model fitness}

For each of the $N=699$ recovery trajectories in which the MCMC sampling satisfied convergence criteria (Section~\ref{sec:convdiag}), we perform posterior predictive checks. That is, we evaluate the location of the observed coral cover within the posterior predictive distribution. For the model to be representing the data accurately we would expect, for any $0 < \beta < 100$, that at least $\beta\%$ of the observations are within the $\beta\%$CrI of the relevant posterior predictive distribution. 

For each observation, we calculate the smallest $\beta$ such that the observation is within $\beta\%$CrI. To do this, samples are generated for the posterior predictive distribution (\eqref{eq:ppd}) and the observed quantiles for each coral cover observation is calculated using,
\begin{equation}
	Q_{\textrm{obs}}^{i,j} = \CondProb{\bvec{X}_s^{(i)}(t_j) \leq \bvec{X}^{(i)}(t_j)}{\dat^{(i)}} = \int_0^{\bvec{X}^{(i)}(t_j)} \CondPDF{\bvec{X}_s^{(i)}(t_j)}{\dat^{(i)}}\, \textrm{d}\bvec{X}_s^{(i)}(t_j), 
\end{equation} 
where $\bvec{X}_s^{(i)}(t_j)$ and $\bvec{X}^{(i)}(t_j)$ are, respectively, the simulated and observed coral cover for the $i$th recovery trajectory at the $j$th time point. Given the observed quantile, then the smallest $\beta$ for the CrI containing the observation is given by $\beta^{i,j} = 100(1-2Q_{\textrm{obs}}^{i,j})$ when $Q_{\textrm{obs}}^{i,j} < 0.5$, and  $\beta^{i,j} = 100(2Q_{\textrm{obs}}^{i,j} - 1)$ otherwise. Hence, we can estimate the proportion of observations that within the posterior predictive $\beta\%$CrI using
\begin{equation}
	\hat{p}_\beta = \frac{\sum_{i=1}^N\sum_{j=1}^{n^{(i)}} \ind{\beta^{i,j} < \beta}}{N\sum_{i=1}^N n^{(i)}}.
\end{equation} 
Here, $\ind{\beta^{i,j} < \beta} = 1$ if $\beta^{i,j} < \beta$ and $\ind{\beta^{i,j} < \beta} = 0$ otherwise. The standard error for the estimate of is given by 
\begin{equation}
	\hat{s}_{\beta} = \sqrt{\frac{\hat{p}_\beta(1-\hat{p}_\beta)}{N\sum_{i=1}^N n^{(i)}}}.
\end{equation} 

\begin{figure}[h]
	\centering
	\includegraphics[width=\linewidth]{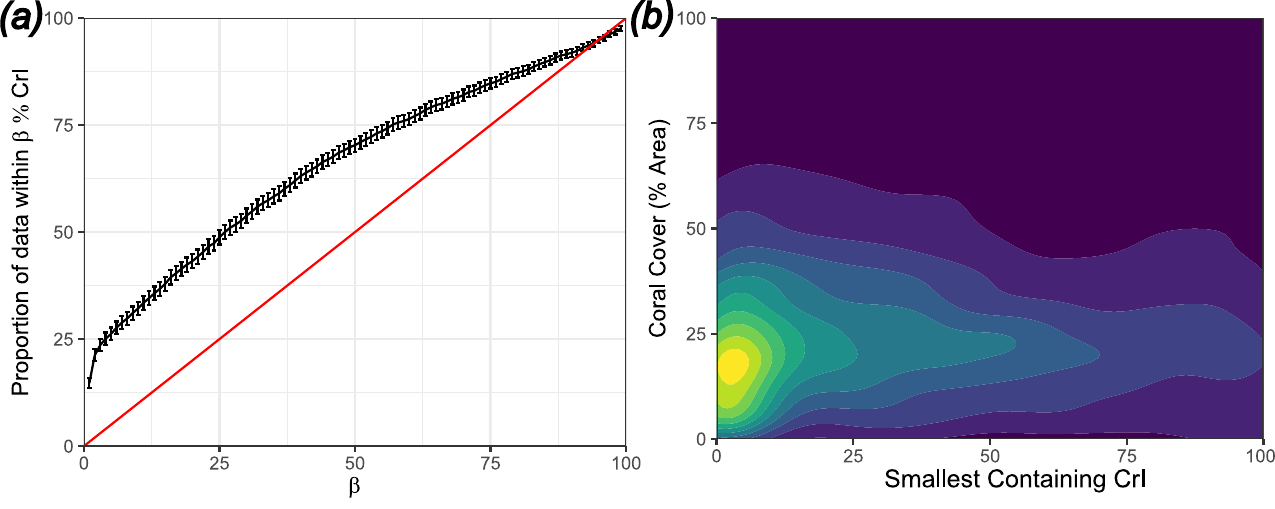}
	\caption{Posterior predictive analysis over the entire GBR data set including $N = 699$ recovery trajectories for which the MCMC sampling converged according to our criteria. (a) The estimated proportion of data that falls within the $\beta\%$CrI, $\hat{p}_\beta$, (solid black lines, with error bars for $95\%$ confidence intervals) for each level of $\beta$ (red line for $\beta = \hat{p}_\beta$). (b) Bivariate density plot of the smallest containing posterior predictive CrI for each coral cover observation (yellow indicates high density and dark blue low density).}
	\label{fig:ppssum}
\end{figure}

Comparing estimates of $\hat{p}_\beta$ with $\beta = 1, 2, \ldots 99$ provides a summary of how well posterior predictive distributions fit the data. As demonstrated in Figure~\ref{fig:ppssum}(a), we observe that for $\beta \in \{1,91\}$, we have $\beta < \hat{p}_\beta \pm 1.96\hat{s}_{\beta}$, indicating the data is well represented by the bulk of the posterior predictive distributions. When $\beta \in \left\{92,93,\ldots,97\right\}$ we observe $\hat{p}_\beta - 1.96\hat{s}_{\beta} < \beta < \hat{p}_\beta +1.96\hat{s}_{\beta}$, indicating the posterior predictive distributions are also tracking the more extreme data points at the expected frequencies. In the extreme tails, $\beta \in \left\{98,99\right\}$ we have $\beta > \hat{p}_\beta \pm 1.96\hat{s}_{\beta}$ and conclude the very extreme variations in the data are under-represented, through the margin of inaccuracy here is still only a few percent. We conclude that overall the performance of the biphasic Richards' model in capturing the variation and dynamics of coral reef recovery is extremely high.

In Figure~\ref{fig:ppssum}(b) a density plot shows the distribution of smallest containing CrIs for each coral cover observation. For coral covers below $25\%$ cover the pattern, the bulk of smallest containing CrIs are well below the $25\%$CrI and the distribution skewed to the right, indicating high posterior probability density for most of the observed data. As coral cover increases, especially above $40\%$ cover, the distribution of smallest containing CrIs is more diffuse. From this we conclude that the majority of the data overdispersion is occurring in regions of higher cover. Importantly, the greatest accuracy in the model is for cover below $30\%$, which is where the majority of the GBR sits. It is also important to emphasise that at higher cover we still observe good fitness, but the data is more variable in the extremes than the model predicts.

\begin{figure}[ht]
	\centering
	\includegraphics[width=\linewidth]{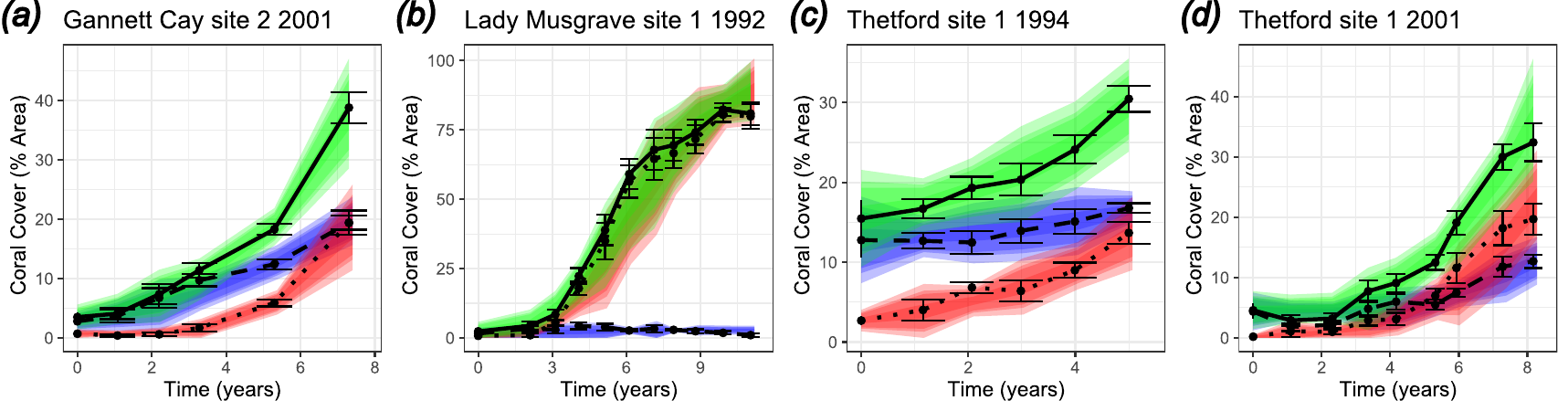}
	\caption{Posterior predictive distributions for example recovery trajectories. (a)--(d) The observed data is given for the total cover for hard coral (solid black lines) broken down into Acroporidae (dotted black lines), and other hard corals (dashed black lines); and model predictions with credible intervals (50\%, 90\%, 95\% and 99\% CrI) as shaded regions for total hard coral (green), Acroporidae (red), and other hard corals (blue).}
	\label{fig:examplepps}
\end{figure}

Figure~\ref{fig:examplepps}(a)--(d) demonstrates example posterior predictive distributions for both the single species model (\eqref{eq:2pglgm}), and the two species model (\eqref{eq:2pglgmms}) that describes the competition for space between fast and slow growing corals. The posterior predictive plots are compared with the observed data in both the single-species and two-species representations for each of the example recovery trajectories: Gannett Cay 2001 (Figure~\ref{fig:examplepps}(a)), Lady Musgrave 1992 (Figure~\ref{fig:examplepps}(b)), Thetford 1994 (Figure~\ref{fig:examplepps}(c)), and Thetford 2001 (Figure~\ref{fig:examplepps}(d)). In each case, there is excellent agreement between the model predictions and the observed data with majority of data points lying within the 95\%CrI of the posterior predictive distribution. 
\FloatBarrier
\subsection{Parameter inference}

\begin{figure}[h!]
	\centering
	\includegraphics[width=\linewidth]{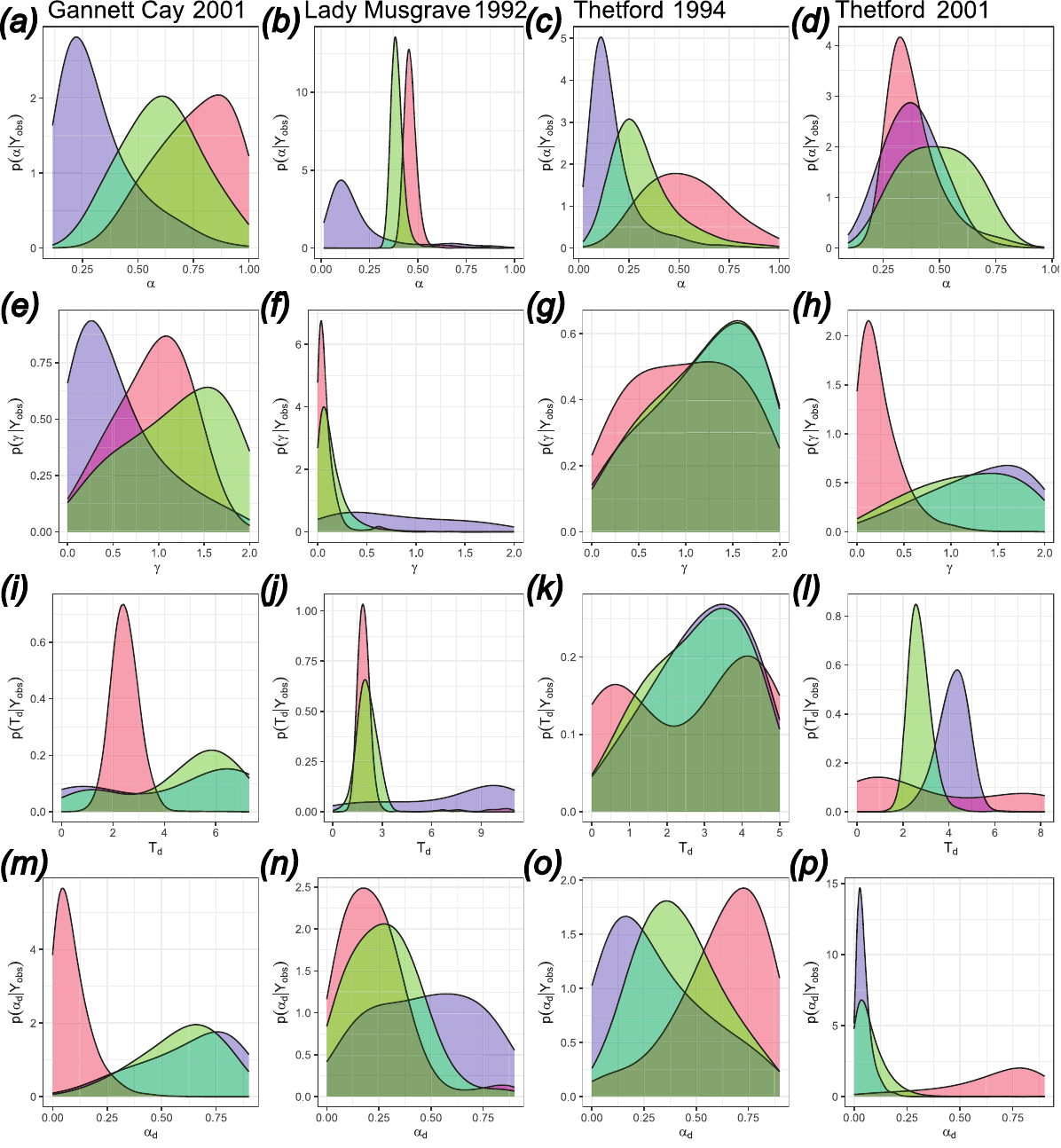}
	\caption{Univariate marginal posterior densities for (a)--(d) the recovery rate $\alpha$, (e)--(h) shape parameter $\gamma$, (i)--(l) change-time $T_d$, and (m)--(p) scale factor $\alpha_d$. Densities are given for total hard coral (green) broken down into \textit{Acroporidae} (red), and other hard corals (blue). Results are shown for the four examples recovery trajectories: (a),(e),(i),(m) Gannett Cay 2001, (b),(f),(j),(n) Lady Musgrave 1992, (c),(g),(k),(o) Thetford 1994, and (d),(h),(l),(p) Thetford 2001.}
	\label{fig:examplemarginals}
\end{figure}

For a given recovery trajectory, marginal posterior densities for the model parameters are estimated directly from the MCMC samples obtained by targeting the full posterior distribution (Sections~\ref{sec:bayesinf} and \ref{sec:convdiag}). These marginal densities can be used to obtain insight into the underlying mechanisms leading to a particular recovery pattern for a recovery trajectory of interest. In Figure~\ref{fig:examplemarginals}, univariate posterior marginal probability densities are shown by coral family group for each of the examples recovery trajectories: Gannett Cay 2001~(Figure~\ref{fig:examplemarginals}(a),(e),(i),(m)), Lady Musgrave 1992 ~(Figure~\ref{fig:examplemarginals}(b),(f),(j),(n)), Thetford 1994~(Figure~\ref{fig:examplemarginals}(c),(g),(k),(o)), and Thetford 1994~(Figure~\ref{fig:examplemarginals}(d),(h),(l),(p)). Among these examples, our model identifies a range of different recovery patterns that could be of interest to reef managers.

Across each of the example trajectories there are some common features in the marginal posterior distributions for the recovery rates (Figure~\ref{fig:examplemarginals}(a)-(d)). Firstly, the two species model tends to indicate the recovery rate for the \textit{Acroporidae} family is greater than the other hard corals, and the recovery rate for the single species model tends to sit between the two sub-groups with the relative weighting between the two subgroups indicative of their relative coral cover contribution (Figure~\ref{fig:examplepps} and Figure~\ref{fig:examplemarginals}(a)-(d)). This is expected since members of the \textit{Acroporidae} family are known to be fast growing by comparison to most other hard corals~\citep{Thompson2010,Ortiz2021}. An interesting deviation from this pattern is in the estimates for the recovery rate for Thetford 2001 (Figure~\ref{fig:examplemarginals}(d)) where the total hard coral estimates are more dispersed than either of the sub-groups. An explanation for this could be the more complex dynamics in the other hard corals with an initial decline before an increase in cover coinciding with the \textit{Acroporidae} corals exceeding the cover of the other hard corals (Figure~\ref{fig:examplepps}(d)). This results in a more complex recovery pattern for the total hard coral cover. 

The shape parameter in the Richards' growth model shows a variety of patterns across the example recovery trajectories (Figure~\ref{fig:examplemarginals}(e)--(h)) with some examples approximating Gompertz growth ($\gamma \to 0$), others approximating logistic growth ($\gamma = 1$), and others tending toward other forms from Richards' growth ($\gamma > 1$). This is not surprising given the diversity in coral species in the GBR. It is worth noting that the posterior distributions for the shape parameter in both Thetford 1994 (Figure~\ref{fig:examplemarginals}(g)) and Thetford 2001 (Figure~\ref{fig:examplemarginals}(h)) are almost identical for the total and other hard corals, which would be expected for the same reef. The difference observed in the posterior for the shape parameter for the \textit{Acroporidae} is likely due to the data containing little information on the sigmoid shape since the recovery for \textit{Acroporidae} is near linear (Figure~\ref{fig:examplepps}(d)). An alternate explanation could be a change in the specific community composition in the \textit{Acroporidae} group at Thetford, we do not explore this further here. We highlight, however, that identifying similarities and differences in the recovery patterns of a specific reef over time is a key use case for coral recovery models in marine ecology and monitoring~\citep{Thompson2020,Ortiz2018}.      

For the parameters that relate to the biphasic recovery, the example recovery trajectories demonstrate the range of different behaviours between different locations and times, and between the single species and two species representations. In the following paragraphs we highlight key features in each example. One important property of our biphasic model (\eqref{eq:2pglgm} and \eqref{eq:2pglgmms}) is that as the scale factor $\alpha_d \to 1$, the change-time becomes structurally non-identifiable. Thus whenever the posterior density for the change-time is very flat, and the posterior density for the scale factor skewed to the left with bulk above $0.6$, then we conclude there is no evidence for biphasic recovery.

In the analysis of Gannett Cay 2001, the single species model indicates little evidence of a biphasic recovery pattern for total hard coral cover. This conclusion is drawn from the relatively flat posterior for the change time in the  (Figure~\ref{fig:examplemarginals}(i)), and the bulk of the scale factor is tending toward larger values (Figure~\ref{fig:examplemarginals}(m)). However, when the data is interpreted with the two species model, there is a strong signal for biphasic recovery patterns are present in the \textit{Acroporidae} group with strong posterior support for change-time between $1$ to $4$ years (Figure~\ref{fig:examplemarginals}(i)) and a small value for the scale factor (Figure~\ref{fig:examplemarginals}(m)). For the other hard corals, the posteriors for the change-time and scale factor are very similar to the single species model with little evidence for biphasic recovery.     

At Lady Musgrave 1992, there is a strong indication for biphasic recovery patterns in total hard coral cover with a change-time of between $1$ to $3$ years (Figure~\ref{fig:examplemarginals}(j)) and scale factor less than $0.6$ (Figure~\ref{fig:examplemarginals}(n)). When using the two species model, it is clear that a biphasic recovery pattern in the \textit{Acroporidae} family, with a change-time of around $1.5$ to $2.5$ years and scale factor less that $0.5$, provides the main driver for the overall biphasic recovery for the total hard coral cover. This is expected at Lady Musgrave where the \textit{Acroporidae} family accounts for almost 100\% of corals (Figure~\ref{fig:examplepps}(b)) \citep{Simpson2022,Murphy2022,Simpson2023}. However, the uncertainty in the biphasic parameters for the small population of other hard corals in the two species model~(Figure~\ref{fig:examplepps}(b)) is reflected in the larger variance in the change-time and scale factor posterior densities for the total hard corals~(Figure~\ref{fig:examplemarginals}(j),(n)).

Analysis for Thetford 1994 and 2001, are consistent with each other in relation to biphasic recovery. In both recovery trajectories, there is evidence for biphasic recovery patterns in the total hard coral cover (Figure~\ref{fig:examplemarginals}(k)--(l),(o)--(p)) (though there is substantially more uncertainty in the results following 1994). Just as in the analysis of Lady Musgrave 1992, one of the species in the two species model seems to be the main driver for biphasic recovery pattern in the total hard coral cover, however, for Thetford it is the other hard corals rather than the \textit{Acroporidae}. We arrive at this conclusion by noting, the change-time and scale factor posteriors densities for the other hard corals follow qualitatively a similar pattern to the equivalent posteriors for total hard corals (Figure~\ref{fig:examplemarginals}(k)--(l),(o)--(p)). The Thetford 2001 recovery trajectory is more informative in relation to the biphasic recovery pattern (Figure~\ref{fig:examplemarginals}(k),(o)), however, both Thetford 1994 and 2001 are qualitatively similar with no evidence for biphasic recovery patterns in the \textit{Acroporidae} family and agreement between other hard corals and total hard corals in support of biphasic recovery patterns (Figure~\ref{fig:examplemarginals}(k)--(l),(o)--(p)). Due to the consistency between distinct trajectories, this would support the theory that this biphasic recovery pattern is a feature of the environment or ecosystem at Thetford reef.

\FloatBarrier

\FloatBarrier
\section{Discussion}
\label{sec:discuss}
In this work, we perform Bayesian parameter inference and posterior predictive checks for $N= 699$ recovery trajectories across the GBR. We find both the single species and two species biphasic Richards' growth model is flexible enough to obtain good model posterior predictive performance. This is impressive in of itself given the range of possible recovery patterns that could arise from the complex interaction of different benthic species. Using four recovery trajectories as exemplars, we highlight the utility of the method for uncertainty quantification in key model parameters related to the Richards' growth shape, recovery rate, and biphasic recovery patterns. Interpretation of these parameters both from a single species and two species viewpoint could support reef management and interventions following significant disturbance events. 

Accurately predicting and interpreting recovery patterns for diverse locations across the GBR is critical for management to support reef recovery, resistance and adaptation into the future~\citep{Ortiz2021,Thompson2020}. Motivated by prior work demonstrating potential widespread deviations from classical Gompertz or logistic recovery models~\citep{Warne2021,Murphy2022,Ortiz2018}, we develop a biphasic extension to the multispecies Richards' growth model. This framework is flexible enough to capture a range of possible sigmoid recovery patterns and biphasic recovery patterns where a period of inhibited recovery occurs in the year immediately following the disturbance event~\citep{Warne2021}. In this paper, we formulate this new model and demonstrate its efficacy in accurately capturing the recovery patterns of nearly every recovery trajectory available within LTMP reef monitoring data, which is spatially and temporally the largest reef monitoring dataset in the world. Therefore, our modelling approach is describing well the recovery patterns observed in the largest and most interconnected coral reef worldwide. Through Bayesian uncertainty quantification we can reliably characterise the combination of parameters that lead to particular recovery behaviours for a reef and provide forecasts of coral recovery following a disturbance event. 

 In this work, we considered the LTMP data from a single species perspective, based on total hard coral cover, and from a two species perspective that separates the fast-growing \textit{Acroporidae} coral families from the other hard corals~\citep{Thompson2010,Thompson2020}. We believe that both models are important for management. Firstly, the single species model enables a macroscopic view of the reef recovery patterns that is useful as a reef health indicator~\citep{Thompson2020} and the notion of a biphasic recovery has a direct interpretation in terms of expected future coral cover that, in turn, can support intervention triage~\citep{Warne2021}. In addition, the single species model is extremely efficient for the purposes of model calibration due to the lower dimensionality of the parameter space and the analytic solution to the biphasic Richards' growth model~\citep{Tjorve2010,Simpson2022}. However, the two species model provides more insight around the potential mechanisms driving any biphasic recoveries that are observed using the single species model. This provides additional information to management authorities regarding the types of interventions that may be applicable. Unfortunately, this additional granularity of interpretation comes with a computational cost as numerical schemes are needed to solve the governing equations and there are more parameters that can lead to slower convergence in the MCMC sampling. Therefore, we recommend a two-stage process for monitoring in which the single species model is used to triage reefs affected by biphasic recovery, then the two species model is introduce to assist in planning interventions. Future work could also explore the use of frequentist methods based on likelihood profiles~\citep{Simpson2023} as a preconditioning step to accelerate the MCMC sampling and reduce the computational burden as proposed by~\citet{Warne2023}. Such improvements would be important if further refinements are considered, such as the inclusion of additional hard coral family divisions or the inclusion of other benthic life forms that compete for space with coral such as macro-algae and sponges~\citep{Thompson2010,Mumby2009}, or the inclusion of spatial hierarchical patterns~\citep{Johns2014}.
 
One limitation to our process is the pre-estimation of maximum coral cover. While this is a standard procedure in the coral monitoring literature~\citep{Thompson2015,Thompson2020}, there is evidence to suggest that these estimates could be overestimates to the effective maximum coral cover~\citep{Cresswell2024}. In addition, it is also mathematically problematic to pre-estimate long-term phenomena, in general. We also note that this pre-estimation of the maximum coral cover is one of the main reasons for the relatively few situations in which we obtain poor model fits or convergence issues with MCMC sampling. Therefore, future work should also consider the inclusion maximum coral cover parameter in the model calibration process. This will be especially important if there is latent competition with other species, like macro-algae, that are not currently explicitly modelled.

We have only considered deterministic models designed to capture the average recovery patterns. As a result, the noise is assumed to be entirely due to the monitoring process~\citep{Jonker2008,Thompson2015}. However, at a colony level, reef growth and competition with other species is a stochastic process~\citep{AlvarezNoriega2023,Bozec2021,McDonald2023}. The net effect of this intrinsic noise could be captured at the reef scale using stochastic differential equation (SDE) models. While the use of SDEs generally requires more complex computational approaches to perform parameter inference~\citep{Cranmer2020, Drovandi2010,Sisson2018,Wang2024,Warne2020,Warne2023}, there are known benefits in using SDE models when intrinsic noise is relevant, as they lead to more accurate parameter inferences and can resolve issue related to  parameter non-identifiability~\citep{Browning2020,Munsky2009}. This could be of considerable benefit, as practical identifiability issues arise in the uncertainty quantification analysis for some parameters (Figure~\ref{fig:examplemarginals}) and is it not feasible to collect more data.   

Prior work by~\citet{Warne2021} demonstrated the importance of taking biphasic recovery patterns in coral reefs into account. Through the application of our Bayesian framework that accounts for this phenomena, the construction of coral reef health indicators that are robust to deviations in standard assumptions could be constructed~\citep{Simpson2022, Thompson2010,Thompson2020}. This new approach enables a more nuanced view of expected reef recovery and provides a modelling basis to investigate the biological, ecological and environmental factors that contribute to biphasic recovery patterns~\citep{Warne2021}. New measures of reef decline can also be established with this approach. For example, tracking the distribution of the change time and scale factor for a given reef over multiple recoveries can identify steady increases in the duration of the slower initial recovery phase that could indicate degradation in reef health. We anticipate our modelling and analysis approach to have direct benefit for the management of the GBR. Finally, through the use of decision-theoretic approaches to sampling design~\citep{Akinlotan2023}, understanding biphasic recovery patterns could enable new monitoring protocols that maximise information for decision makers.

\paragraph{Acknowledgements}
This work was supported through the Australian Institute of Marine Science (AIMS) and the Centre for Data Science at the Queensland University of Technology. Computational resources were provided by the eResearch Office, Queensland University of Technology.

\end{document}